\documentclass[prd,twocolumn,amsmath,amssymb]{revtex4}

\usepackage{graphicx}
\usepackage{dcolumn}% Align table columns on decimal point
\usepackage{bm}% bold math

\def\n3nu{$\it n \rightarrow \nu \nu \bar{\nu} \ $}
\def\nnbari{$ \it n \rightarrow \bar{n} \ $}
\def\taun3nu{$\rm \tau_{n3\nu} \ $} 

\begin{document}

\title{Signatures of Nucleon Disappearance \\in Large Underground Detectors \\}

\author{Yuri Kamyshkov}
 \email{kamyshkov@utk.edu}
 \affiliation{Department of Physics, University of Tennessee, Knoxville, TN 37996}

\author{Edwin Kolbe}
 \email{kolbe@quasar.physik.unibas.ch}
 \affiliation{Physics Division, Oak Ridge National Laboratory, Oak Ridge, TN 37831}

\begin{abstract}
For neutrons bound inside nuclei, baryon instability can manifest 
itself as a decay into undetectable particles (e.g., \n3nu), i.e., as
a disappearance of a neutron from its nuclear state. If electric charge 
is conserved, a similar disappearance is impossible for a proton. 
The existing experimental lifetime limit for neutron disappearance is 4-7 
orders of magnitude lower than the lifetime limits with detectable nucleon 
decay products in the final state \cite{PDG2002}. 
In this paper we calculated the spectrum of nuclear de-excitations 
that would result from the disappearance of a neutron or two neutrons 
from $^{12}$C. We found that some de-excitation modes 
have signatures that are advantageous for detection in the modern 
high-mass, low-background, and low-threshold underground detectors, where
neutron disappearance would result in a characteristic sequence 
of time- and space-correlated events. Thus, in the KamLAND detector \cite{Kamland}, 
a time-correlated triple coincidence of a prompt signal, a captured neutron, 
and a $\beta^{+}$ decay of the residual nucleus, all originating from the 
same point in the detector, will be a unique signal of neutron 
disappearance allowing searches for baryon instability with sensitivity 
3-4 orders of magnitude beyond the present experimental limits.\\

\noindent
PACS number(s):~24.80.+y, 13.30.Ce, 14.20.-c, 24.10.Pa
\end{abstract}

\maketitle

\section{\label{sect1}Introduction}

Baryon asymmetry of the universe \cite{Sakharov67} and the idea of unification 
of particle and forces \cite{Salam73,Georgi74} are the two global concepts 
that motivated the nucleon decay experimental searches \cite{Goldhaber96} for 
more than two decades. The particular modes of nucleon decay are not 
known a priori, although some are preferred by theoretical models 
\cite{Salam73,Georgi74,Babu2000,Mohapatra2001}. The results from several 
generations of nucleon decay search experiments \cite{PDG2002}, including the 
most spectacular recent Super-Kamiokande results \cite{SuperK}, have 
set stringent limits on many exclusive nucleon decay modes. For example, the 
lifetime limit for the decay mode $p \rightarrow e^{+} + \pi^{0}$, established by the 
Super-K Collaboration, reaches $\tau_{p} \geq 5 \cdot 10^{33}$ years 
\cite{SuperK01}. However, the mode-independent nucleon lifetime limit \cite{PDG2002} 
is established experimentally only at the level of $\geq 1.6 \cdot 10^{25}$ years 
\cite{Evans77}, which is more than 8 orders of magnitude poorer than the 
previous limit. That indicates that experimental searches for certain decay modes 
need to be extended.

The Particle Data Group identifies 75 possible modes of nucleon decay 
(baryon instability) \cite{PDG2002} that respect the conservation laws 
of electric charge, energy-momentum, and angular momentum. The lifetime 
limits for most of these modes lie above $10^{30}$ years, but there are 
a few significant exceptions. These are the modes corresponding to the decay 
of one neutron or two neutrons into neutrinos: $n \rightarrow 3 \nu$, 
$n \rightarrow 5 \nu$, and $nn \rightarrow 2 \nu$. Lifetime limits for 
these modes are several orders of magnitude lower \cite{PDG2002}. Generally 
speaking, these modes can be interpreted as disappearance of one or two 
neutrons from their intra-nucleus state into any undetectable particles. 
Following the spirit of the paper of Evans and Steinberg \cite{Evans77} 
and using the assumptions of the conservation laws mentioned above, one 
could think that the present ``mode-independent'' nucleon decay limit \cite{PDG2002} 
is determined by these neutron disappearance modes. We believe that 
experimental searches for these modes and the improvement of mode-independent 
nucleon lifetime limit are as fundamentally important as 
searches for particular exclusive nucleon decay modes with super-giant 
detectors. 

Moreover, decays like $N \rightarrow lepton + lepton + antilepton$ 
arise in certain unification schemes \cite{Salam73,Pati84} as a result 
of $(B-L)$ non-conserving interactions that might be more natural for 
the explanation of baryon asymmetry of universe than $(B-L)$-conserving 
transitions. The question of whether $(B-L)$ is violated in nature and 
at what energy scale can be answered only by experiments \cite{Yuri99}. 
Nucleon decay into five and more leptons was theoretically considered in 
\cite{Salam73}; dinucleon decay into two leptons in \cite{Feinberg78}.

In this paper we demonstrate that the disappearance of a neutron 
(or two neutrons) from its intra-nucleus state creates subsequent 
nuclear de-excitations that, for a few particular de-excitation modes,
would lead to a unique experimental signature: {\em a chain of time- and 
space-correlated events}. Such a signature can be observed in the modern 
large-mass, low-background, and low-threshold detectors, e.g., in KamLAND 
\cite{Kamland}. We estimate that the sensitivity of a search for such events 
in the KamLAND detector will result in improvement of one-neutron and 
two-neutron disappearance limits by 3-4 orders of magnitude beyond 
the existing lifetime limits.

The idea that nucleon decay in the nucleus can be studied by measuring 
particles and/or $\gamma$ rays accompanied by de-excitation of 
the nucleon hole was, to our knowledge, first considered 
by Y. Totsuka \cite{Totsuka86} and H. Ejiri \cite{Ejiri93} 
and used by the Kamiokande Collaboration \cite{Suzuki93} as 
discussed below. Also, recently the DAMA Collaboration \cite{Dama00} has searched 
in 6.5 kg of a liquid $^{129}$Xe scintillator detector operating in Gran Sasso 
National Laboratory for the decay products of the residual radioactive daughter 
nuclei left after the disappearance of one or two nucleons from a $^{129}$Xe 
nucleus and set new limits for two-nucleon disappearance.

In section \ref{sect2} of this paper, we discuss the existing experimental 
limits for $n \rightarrow 3 \nu$, $nn \rightarrow 2 \nu$, and, in general, for disappearance
of one or two neutrons. Our calculations for de-excitation branching ratios 
in intra-nucleus disappearance of one neutron from $^{12}$C are discussed in 
section \ref{sect3}; for comparison, in section \ref{Osect}, we also calculated 
the disappearance of a neutron from an $^{16}$O nucleus; section \ref{sect4} deals 
with the calculations of de-excitation modes following the two-neutron 
disappearance from $^{12}$C. The possible observation of some particular 
de-excitation modes followed by the radioactive decay of daughter nuclei in the 
KamLAND detector, as well as the possible sources of background, are discussed in 
section \ref{discuss}. Our conclusions are presented in section \ref{conclude}.

\section{\label{sect2}Existing limits for neutron disappearance}

Experimental limits have been set for $n \rightarrow 3 \nu$ decay
in underground experiments \cite{Learned79,Berger91} assuming that 
the whole Earth is the source of decaying neutrons. The underground
detectors in this case measure the specific energy spectrum above the 
background level of atmospheric neutrino signals, resulting from 
decay-neutrinos interactions with the material of the detector.  
The best limit set from the analysis of the muon events induced by 
charge-current $\nu_{\mu}$ interactions in the detector \cite{Learned79} 
was $\tau (n \rightarrow 3 \nu_{\mu}) > 5 \cdot 10^{26}$ years. The best 
limit from $\nu_{e}$ interactions was set in \cite{Berger91} as 
$\tau (n \rightarrow 3 \nu_{e}) \geq 3 \cdot 10^{25}$ years. 
Due to the detection method, these limits are only valid for the specific 
decay of a neutron into particular types of neutrinos and cannot generally 
be applied to the neutron disappearance.

A very inspirational approach has been used by the Kamiokande Collaboration 
\cite{Suzuki93} who set the limit for $n \rightarrow \nu \nu \bar{\nu}$ 
decay of $\tau \geq 4.9 \cdot 10^{26}$ years, independent of the type of 
neutrinos emitted in the neutron decay. By virtue of the experimental method, 
this limit is also relevant for intra-nucleus neutron disappearance. 
The method is based on the detection of $\gamma$-rays, which could be residual 
de-excitation products of a highly excited $^{15}$O nucleus left after the 
disappearance of a neutron from $^{16}$O in the low-background energy range 
between 19 and 50 MeV. Because of the similarity to our approach, we are 
discussing this method below in more detail. In this regard we also refer to
the paper by H.~Ejiri \cite{Ejiri93}.

When a neutron disappears from the oxygen nucleus, the residual $^{15}$O 
is often left in an excited state. In the nuclear shell model the 
excitation level is determined by the level that a neutron has inside 
the $^{16}$O and by the width of this energy level (spreading width). 
In the $^{16}$O nucleus, two neutrons occupy the lowest $s_{1/2}$ level, 
4 neutrons fill the next $p_{3/2}$ shell, and the remaining 2 neutrons 
populate the highest $p_{1/2}$ state. The disappearance of a neutron from 
the $p_{1/2}$ state would result in a minimal excitation where the residual 
$^{15}$O with high probability will be left in the ground state and 
the difference of the binding energy will be taken care of by undetected neutron 
decay products. The disappearance of a neutron from a lower $p_{3/2}$ state will 
create a hole that will result in a restructuring of the residual nucleus with 
possible emission of $\gamma$ radiation. Thus, it was estimated in \cite{Suzuki93} 
that de-excitation of the $p_{3/2}$ neutron hole with a probability of $\sim$ 44\% 
will produce  6.18 MeV $\gamma$-rays. The remaining part the $^{15}$O 
nucleus will again be left in the ground state. When a disappearing neutron 
produces a hole in the $s_{1/2}$ shell, the excitation energy is large enough to 
exceed the separation energy for protons and neutrons in $^{15}$O. In this case
de-excitation proceeds mainly by emission of a proton, neutron, $\alpha$, 
and, to a smaller extent, by emission of $\gamma$s. Since the Kamiokande 
water-\v{C}herenkov detector was not sensitive to low-energy protons, 
neutrons, $\alpha$s, and had a threshold for $\gamma$ and electron 
detection of 7.5 MeV, the authors of paper \cite{Suzuki93} concentrated 
only on the probability of de-excitation of the $s_{1/2}$-state hole via 
emission of $\gamma$s within an energy range of 19-50 MeV, where the experimental 
background was very small. This resulted in a rather small de-excitation 
branching ratio (Br) estimated in \cite{Suzuki93} as $(0.27 - 1.04) \cdot 10^{-4}$. 
Due to this small branching ratio, the measured 
$\tau /Br = 1.8 \cdot 10^{31}$ years in the Kamiokande detector has resulted 
in $\tau_{n \rightarrow 3 \nu}$ disappearance limit of only 
$\geq 5 \cdot 10^{26}$ years. 
  
For completeness, we need to mention that the Particle Data Group listings 
\cite{PDG2002} also include the $n \rightarrow 3 \nu$ and $n \rightarrow 5 \nu$
lifetime limits of the J.-F. Glicenstein analysis \cite{Glicenstein97} 
$\tau (n \rightarrow 3 \nu_{X}) \geq 2.3 \cdot 10^{27}$ years and
$\tau (n \rightarrow 5 \nu_{X}) \geq 1.7 \cdot 10^{27}$ years, based 
on the Kamiokande experimental data and on the idea that the 
disappearance of the neutron's magnetic moment should produce radiation.

Existing lifetime experimental limits for $nn \rightarrow 2 \nu$ are even less stringent.
The following two limits were determined by the Fr\'{e}jus Collaboration 
\cite{Berger91}: $\tau (nn \rightarrow 2 \nu_{e}) \geq 1.2 \cdot 10^{25}$ 
years and $\tau (nn \rightarrow 2 \nu_{\mu}) \geq 6.0 \cdot 10^{24}$ years.
These limits assume two-neutron decays into neutrinos and antineutrinos of 
{\it e} and {\it $\mu$} flavors rather than the more general intra-nucleus 
disappearance of two neutrons. Recently the DAMA Collaboration \cite{Dama00}, 
looking for the de-excitation products of the residual radioactive daughter 
nuclei left after the two-neutron disappearance from a $^{129}$Xe, has set the
lower limit for $nn \rightarrow 2 \nu$ (more generally for two-neutron disappearance)
as $\tau (nn \rightarrow 2 \nu) \geq 1.2 \cdot 10^{25}$ years. 

\section{\label{sect3}Neutron Disappearance from $^{12}$C}

\begin{figure}[!b]
\includegraphics[width=3in,keepaspectratio]{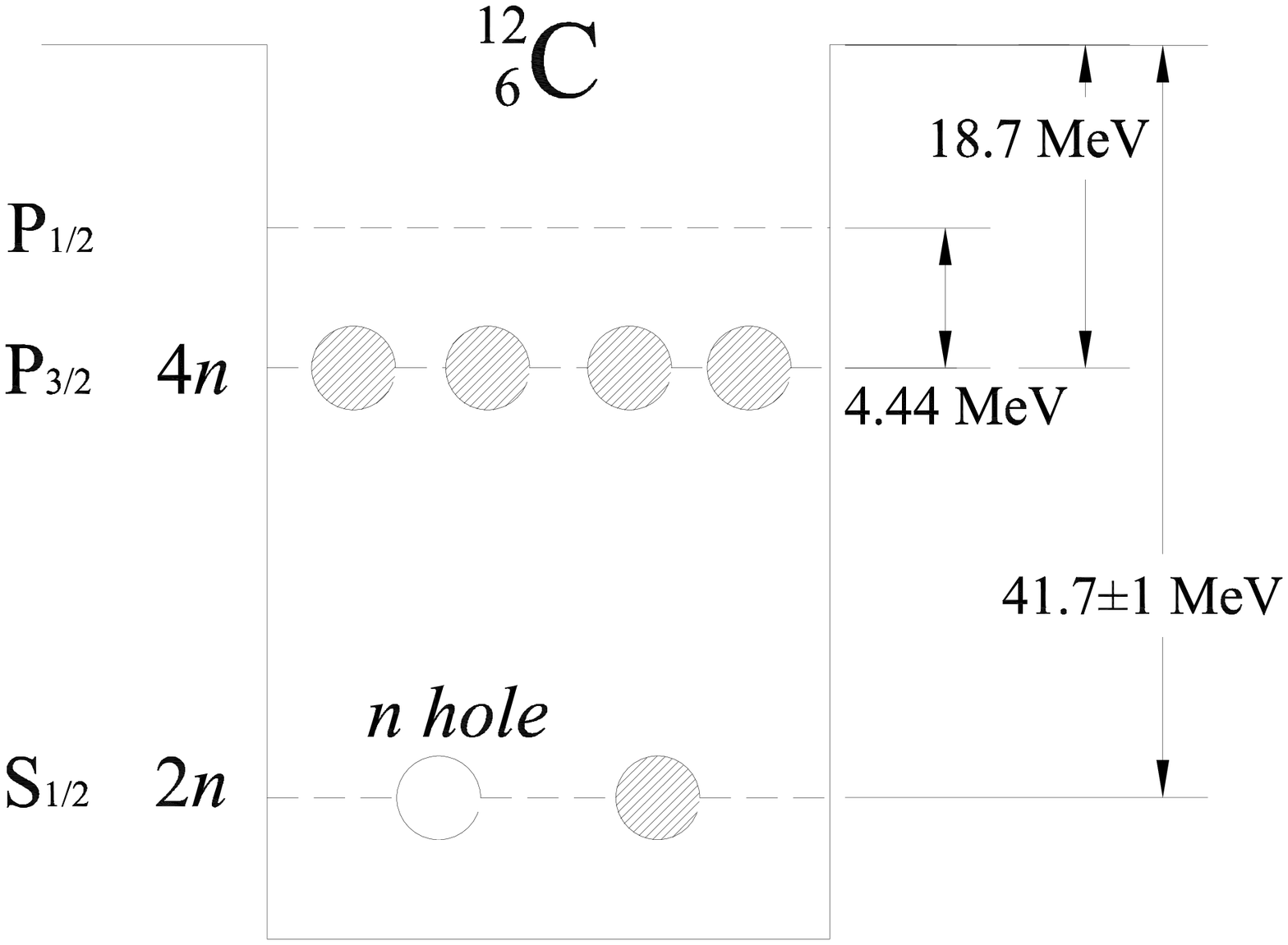}
\caption{\label{fig1} Occupation of energy levels by neutrons 
                      for the $^{12}$C ground state in a simple 
                      shell model picture. One neutron is shown 
                      as disappeared from $s_{1/2}$ level.}
\end{figure}

Large liquid scintillator detectors like KamLAND \cite{Kamland} and 
Borexino \cite{Borexino} contain $^{12}$C as a major component of the 
detector material. In the simple nuclear shell-model picture the 
$^{12}$C ground state is described as a closed-shell nucleus i.e., 
for both protons and neutrons the lowest-lying $s_{1/2}$ and $p_{3/2}$ 
shells are completely filled, while all higher (sub)shells are empty. 
This occupation of the nuclear energy levels by neutrons is sketched in 
Figure \ref{fig1}. Thus, most of the neutrons available in liquid 
scintillator detectors for the baryon instability search are the neutrons
occupying $s_{1/2}$ and $p_{3/2}$ energy levels in $^{12}$C.
 
\subsection{Disappearance from the $p_{3/2}$ shell}

If, in the simple shell-model picture, a neutron disappears from the 
$p_{3/2}$ shell in $^{12}$C, the residual $^{11}$C nucleus is always 
left in the ground state.   
However, in more sophisticated shell-model calculations, it has been shown
that the simple closed-shell picture of the $^{12}$C ground state 
has to be remedied. 
Due to correlations, especially pairing effects, the $p_{1/2}$ shell 
is also partially filled with protons and neutrons from the $p_{3/2}$ level
lying just 4.44 MeV below (Fig.\ref{fig1}).
Various shell-model calculations \cite{ShellMods} give partial occupation 
numbers of the $p_{1/2}$ shell for both neutrons and protons in the range 
0.75$-$0.8, corresponding to an average filling of the $p_{3/2}$ shell 
between 3.2 and 3.25.
If we interpret the partial occupations by saying that approximately 
60 percent of the time the $p_{3/2}$ shell is completely filled, while 
roughly 40 percent of the time two neutrons (the same for a proton pair) 
are in the $p_{1/2}$ shell, then there is a finite probability that a
neutron disappearing from the $p_{3/2}$ state will leave the
residual $^{11}$C nucleus in an excited state with spin and parity 
$J^{\pi}=1/2^{-}$. 
There is only one state in $^{11}$C with $J^{\pi}=1/2^{-}$ 
below 10 MeV~\cite{Aj90}, which is the first excited state at 2.0 MeV.
Therefore, we assume that this is the only possible final state when 
a $p_{3/2}$ neutron disappears, while at the same time a pair of neutrons is 
in the $p_{1/2}$ shell, i.e., we assume the corresponding spectroscopic 
factor to be 1.
Then the probability per P-shell neutron for this event is given by the 
product of 40 percent and 1/2, that is 20 percent, with an adopted error 
of $\pm 5$ \%. The first excited state in $^{11}$C will promptly de-excite 
by emission of a $\gamma$ with an energy of 2.0 MeV, and the ground state 
of $^{11}$C will subsequently undergo a $\beta^{+}$ transition 
(the half-lifetime of $^{11}$C is 20.4 min and Q$_{EC}$ = 1.98 MeV). 

Thus, the disappearance of a neutron from the $p_{3/2}$ state will result, 
with a branching of 20$\pm$5\% per each of four P-shell neutrons,
in two time-correlated events: 
emission of monoenergetic 2.0 MeV $\gamma$ and $\beta^{+}$ emission with 
detectable energy in the range 1.02--1.98 MeV.

\subsection{Disappearance from the $s_{1/2}$ shell}

A more complicated and rich picture emerges if a neutron 
disappears from the $s_{1/2}$ state in $^{12}$C. Excitation 
energy of the residual $^{11}$C$^{*}$ in this case is very 
high; it exceeds the separation energy for proton, neutron, 
and $\alpha$ particles in $^{11}$C, making them the primary  
emission products of the highly excited nucleus. Qualitatively,
emission of a proton should be more frequent than emission of 
a neutron, since $^{11}$C is a proton-rich nucleus. Primary 
emission might not be sufficient for removing all the 
excitation energy and therefore might be followed by secondary
de-excitation. Nuclear model calculations are needed in order 
to estimate these particular de-excitation branching ratios.

The residual $^{11}$C nucleus is left in a highly excited
state with (assuming conservation of angular momentum and parity) 
total spin and parity $J^{\pi} = \frac{1}{2}^{+}$.
The excitation energy $E^{*}$ of this state can be calculated 
by the difference between 
the binding energy of the $s_{1/2}$ neutron level in $^{12}$C and the
neutron separation energy of $^{12}$C.
While the latter is well known to be $S_n = 18.72$~MeV, no precise 
experimental or theoretical values exist for the binding energy of the 
$s_{1/2}$ neutron level. 
In the case of the $s_{1/2}$ proton level in $^{12}$C 
the binding energy and the width have been measured in 
several electron scattering experiments $^{12}$C($e,e^\prime$p) 
\cite{C12eep} and were determined to be $E_{1S}^p = 39 \pm 1$~MeV and 
$\Gamma^p_{1S} = 12 \pm 3$~MeV, respectively. Correcting $E_{1S}^p$ for 
the Coulomb shift of $\approx 2.7$~MeV for protons in $^{12}$C gives 
us $E_{1S}^n$=41.7 $\pm$ 1~MeV for the binding energy of the $s_{1/2}$ 
neutron level in $^{12}$C and, subtracting the neutron separation energy, 
we get the value $E^{*}=23 \pm 1$~MeV for the excitation energy in $^{11}$C. 
Determination of width of the excited state caused by the disappearance 
of $s_{1/2}$ nucleon is not that straightforward. In terms of nuclear correlations 
that are responsible for the width of the excited state, the disappearance of 
nucleon a priori is not the same process as the knocking-out of a nucleon, 
since in the latter case a strongly-interacting particle remains present in the 
final state inside the nuclear matter. Different approaches are possible here. 
Theoretical estimates for the spreading width of the $s_{1/2}$ hole in infinite 
nuclear matter \cite{Muether_private} gives a value of $\sim$6 MeV. As another 
extreme one can use the spreading width of $s_{1/2}$ state as determined in 
proton knock-out exeriments \cite{C12eep}. We adopt what we think is a reasonable 
compromise here and assumed that the excited level at $E^{*}=23 \pm 1$~MeV 
in $^{11}$C has a Lorentzian width of $\Gamma = 7$~MeV. At the end of the 
section we will show that our main results do not sensitively depend 
on the values adopted for $E^{*}$ and $\Gamma$ in the whole range of variations
mentioned above.

In order to determine the de-excitation modes of the excited $^{11}$C nucleus,
i.e., the branching ratios in the various particle emission channels and 
the energy spectra of the emitted particles, we used the statistical model 
of compound nuclear reactions (Hauser-Feshbach) \cite{StatModel}. 
Within this model the branching ratio for the decay $A^*(c)A_c$
of an excited state of a nucleus $A$ with energy $E^*$ and spin and 
parity $J^{\pi}$ to a final nucleus $A_c$ by emission
of a particle $c$ is given by the ratio of the transmission coefficients:
\begin{eqnarray}
  Br_c (E^*,J^\pi) & = & \frac{T_c(E^*,J^\pi)}{T_{tot}(E^*,J^\pi)}
\end{eqnarray} 
Here $T_{tot}$ is simply the sum of the transmission coefficients of all 
decay channels open at $E^*$:
\begin{eqnarray}
  T_{tot}(E^*,J^\pi) & = & \displaystyle\sum \limits_{c} T_{c}(E^*,J^\pi) 
\end{eqnarray} 
which guarantees that the sum of the branching ratios is normalized to one.
Each of the transmission coefficients $T_{c}$ is obtained by adding the 
contributions from all allowed final states in the residual nucleus:
\begin{eqnarray}
 \label{Tcoff}
  T_{c}(E^*,J^\pi) & = & 
   \displaystyle\sum \limits_{\mu=0}^{N_c} 
    T^\mu_{c}(E^*,J^\pi, E_c^\mu,J_c^\mu,\pi_c^\mu) + \nonumber \\
   & & \hspace*{-8em} \displaystyle\int \limits_{E_c^{N_c}}^{E^*-S_c} 
    \displaystyle\sum \limits_{J_c,\pi_c} T_{c}(E^*,J^\pi,E_c,J_c,\pi_c)
     \rho(E_c,J_c,\pi_c) dE_c 
\end{eqnarray} 
In this sum the decays to the ground state ($\mu = 0$) and all 
experimentally known excited levels ($\mu = 1,2,\ldots$,$N_c$) in the 
daughter-nucleus are explicitely taken into account.
For excitation energies higher than the energy $E_c^{N}$ of the last
known level the sum is changed to an integration over the level density
$\rho$ up to the highest energetically allowed state at energy $E^*-S_c$,
where $S_c$ is the channel separation energy for the ejected particle $c$.
It can be seen, that the important ingredients of 
statistical model calculations are the particle and $\gamma$-transmission  
coefficients $T$ and the level density of the excited states.
Therefore, the quality of the calculated results within the
statistical model is determined by the accuracy with which these
components can be evaluated.  

The branching ratios and particle energy spectra we present 
in the following have been calculated with the 
statistical model code SMOKER \cite{Smoker}, which applies realistic
optical potentials (reproducing nicely experimental data) to
determine the transmission coefficients and a sophisticated
description for the level density (by a combination of 
a constant temperature formula and a back-shifted Fermi gas). 
Starting from an excited state with well-defined energy, angular 
momentum, and parity, the SMOKER program calculates the branching ratios for 
transitions via photon, neutron, proton, and $\alpha$ emission, which are the 
dominant de-excitation channels. 
Note that within SMOKER the transmission coefficients in Eq.~(\ref{Tcoff})
describing the transition from the excited state ($E^*,J^\pi$)
to the state ($E_c^\mu,J_c^\mu,\pi_c^\mu$) in the daughter nucleus
are correctly obtained by summing up the contributions from all
quantum mechanically allowed partial waves: 

\begin{eqnarray}
T^\mu_{c}(E^*,J^\pi, E_c^\mu,J_c^\mu,\pi_c^\mu) & = & \nonumber \\ 
& & \hspace*{-10em} \displaystyle\sum \limits_{l=|J-s|}^{J+s} \;
\displaystyle\sum \limits_{s=|J_c^\mu-j_c|}^{J_c^\mu+j_c}
\: T_{c_{ls}}(E^* - S_c - E_c^\mu) 
\end{eqnarray} 

\begin{figure*}[t]
\includegraphics[angle=90,width=5.8 in,bb=51 55  517 771,clip]{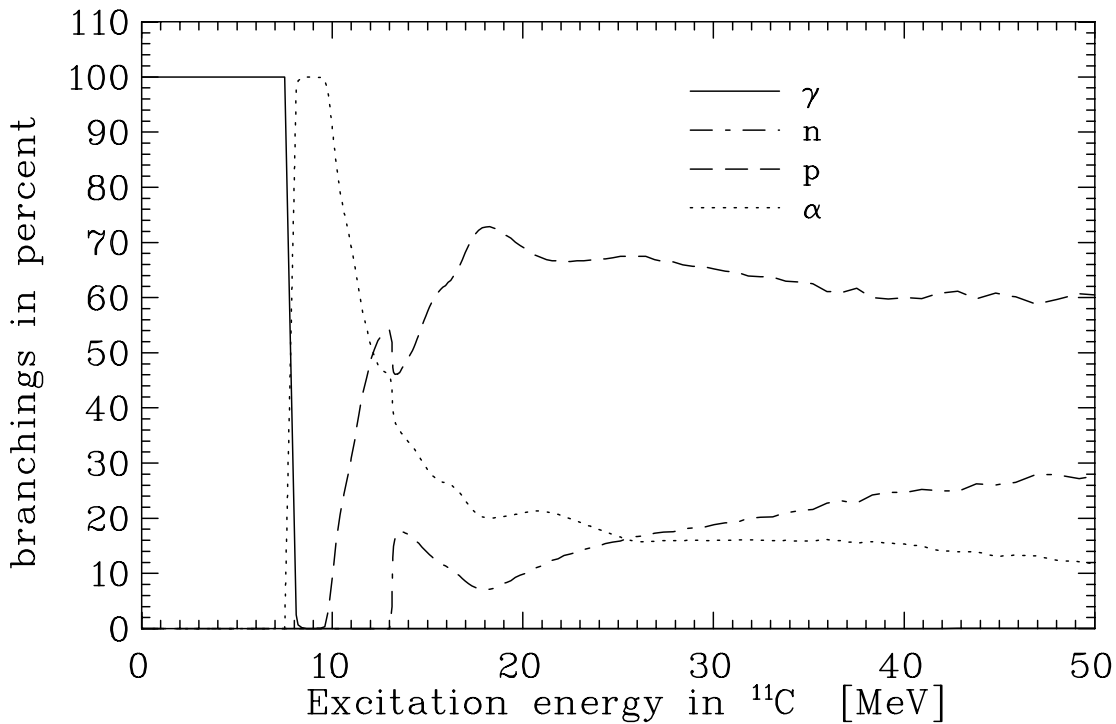}
\caption{\label{fig2}Energy dependence of de-excitation branching 
   of a $J^{\pi} = \frac{1}{2}^{+}$  $^{11}$C$^{*}$ state by emission of 
   a $\gamma$, neutron, proton, or $\alpha$ particle as built into the
   statistical model (SMOKER code).}
\end{figure*}

Here angular momentum $\vec{l}$ and the channel spin 
$\vec{s} = \vec{j}_c + \vec{J}_c^\mu$ couple to 
$\vec{J} = \vec{l} + \vec{s}$.
For the physical quantities typically needed in statistical model
calculations (e.g., nuclear masses, level density parameters),  
SMOKER takes the experimental values, if they exist. Otherwise, 
theoretical models are applied. 
This is illustrated in Figure~\ref{fig2}, where the branching ratios
for the decay of an excited state of $^{11}$C$^{*}$ with spin and parity 
$J^{\pi} = \frac{1}{2}^{+}$ into the photon, neutron, proton, and $\alpha$ 
channels are shown as functions of excitation energy.
Note that all particle decay channels open at the correct thresholds  
($S_\alpha = 7.54$~MeV, $S_p = 8.69$~MeV, $S_n = 13.12$~MeV) because
the experimental masses of the involved nuclei have been used.
With increasing excitation energy, as soon as a hadronic decay channel
opens, the branching for $\gamma$-decay (electro-magnetic interaction)
becomes negligible. While the contribution to the neutron channel
rises immediately after the threshold, the branching into the proton 
channel needs 2-3 MeV above the threshold to become open because the 
Coulomb barrier has to be overcome.

If particle emission decay leads to an excited level of 
the residual nucleus, we again calculate the branching ratios for 
the subsequent emission with the statistical model (iteration of SMOKER).
By keeping track of the energies of the ejected particles and
photons during the cascade, and by weighting them with the corresponding
branching ratios, the energy spectra for emitted neutrons, protons, 
$\alpha$s, and $\gamma$s can be determined. Since the amount of 
excitation energy in the system is significantly reduced with every 
particle emission and eventually drops below the particle emission 
thresholds, we found that the cascade could practically be terminated 
after two emission steps, i.e., the residual daughter nuclei after two 
iterations were assumed to be in their ground state. 
 
The SMOKER code has been applied extensively to many astrophysical
problems, mainly where nuclei with mass number $A>20$ are involved.
In such a case the requirement of a high level density, which is one of 
the main assumptions of the statistical model, is usually met, and the 
calculated cross sections are generally in good agreement with the experimental 
data \cite{Smoker}. For the light nuclei considered here, an application of 
the statistical method might seem problematic because the level 
density in light nuclei can be small, and in many reactions with light 
nuclei, significant direct contributions have been found.
However, by studying the de-excitations of $^{12}$C and $^{16}$O nuclei 
that had been excited by inelastic neutrino scattering, it was found 
that the branching ratios for decay into the proton and neutron channels, 
calculated within the statistical model, were in very good agreement with 
the branching ratios obtained from a continuum random phase approximation
~\cite{mythesis}. That means that at certain conditions, direct contributions 
can be effectively accounted for in the statistical model.
Although at specific excitation energies, the branchings 
calculated within both models deviated by a factor of 2, the proton and 
neutron branchings averaged over various excitation spectra corresponding 
to different neutrino energies were found to match within 1-17\%. 
As for our calculations in this paper, due to the spreading width of the 
excited state, the branching ratios from the statistical model are also 
averaged over a range of excitation energies. 
Therefore, we conservatively adopt the relative error in the branchings 
for one- and two-step decays to be 20\% and 30\%, respectively.

We now turn to the results of our calculations of the final states of 
nuclear de-excitations following the disappearance of a neutron from the $s_{1/2}$ 
shell in $^{12}$C. Results of these calculations are shown in Table~\ref{tab1}. 
They were obtained for an excited $J^{\pi} = \frac{1}{2}^{+}$ level at 
$E^{*}=23$~MeV in $^{11}$C with an associated Lorentzian width of 
$\Gamma$=7~MeV.

\begin{table}[h!t]
\caption{ \label{tab1}Branching ratios for $^{11}$C$^*$ de-excitation modes
         after a neutron disappears from the $s_{1/2}$-state in 
         $^{12}$C and experimental signature (number of time-correlated hits)
         for observation of these modes in a large liquid scintillator 
         detector. For the final states marked with $^{*}$, due to the very 
         large lifetime of the daughter nuclei, the experimental signature is 
         assumed to be a single hit.}
 \begin{center}
   \begin{tabular}{|l|l|r|c|} \hline \hline
     Decay & Daughter & Mode  & Exp.   \\ 
     mode  & (decay, T$_{1/2}$ or $\Gamma$, Q$_{EC}$ or Q$_{\beta ^{-}}$)  & \%    & sign.  \\ 
           &          &       & (hits) \\ \hline\hline
    $^{11}$C($\gamma$)        & $^{11}$C$_{\rm gs}$($\beta^+$; 20.4 m, 1.98 MeV) 
       &  0.7  &  2 \\ 
    $^{11}$C(n$\ldots$)       & $\downarrow$       
       & 13.8  &  $\downarrow$ \\
    $^{11}$C(p$\ldots$)       & $\downarrow$       
       & 64.4  &  $\downarrow$ \\
    $^{11}$C($\alpha \ldots$) & $\downarrow$ 
       & 21.1  &  $\downarrow$ \\ \hline\hline
    $^{11}$C(n)               & $^{10}$C$_{\rm gs}$($\beta^+$; 19.3 s, 3.65 MeV)
       &  3.0  &  3 \\ 
    $^{11}$C(n,$\gamma$)      & $^{10}$C$_{\rm gs}$($\beta^+$; 19.3 s, 3.65 MeV)
       &  2.8  &  3 \\ 
    $^{11}$C(n,n)             & $^{9}$C($\beta^+$; 0.127 s, 16.5 MeV)     
       &  0.06  &  4 \\ 
    $^{11}$C(n,p)             & $^{9}$B(p+2$\alpha$; 0.54 keV, 1.07 MeV)     
       &  5.7  &  2 \\ 
    $^{11}$C(n,$\alpha$)      & $^{6}$Be(2p+$\alpha$; 92 keV, 4.3 MeV) 
       &  2.2  &  2 \\ \hline
    $^{11}$C(p)               & $^{10}$B(stable)
       &  2.9  &  1 \\ 
    $^{11}$C(p,$\gamma$)      & $^{10}$B(stable)
       & 19.0  &  1 \\ 
    $^{11}$C(p,n)             & $^{9}$B(p+2$\alpha$; 0.54 keV, 1.07 MeV)     
       &  1.4  &  2 \\ 
    $^{11}$C(p,p)             & $^{9}$Be(stable)    
       &  7.2  &  1 \\ 
    $^{11}$C(p,$\alpha$)      & $^{6}$Li(stable) 
       & 33.9  &  1 \\ \hline
    $^{11}$C($\alpha$)        & $^{7}$Be($\beta^+$; 53.3d, 0.86 MeV)
       &  5.2  &  1$^{*}$ \\ 
    $^{11}$C($\alpha,\gamma$) & $^{7}$Be($\beta^+$; 53.3d, 0.86 MeV)
       &  4.2  &  1$^{*}$ \\ 
    $^{11}$C($\alpha$,n)      & $^{6}$Be(2p+$\alpha$; 92 keV, 4.3 MeV)     
       &  0.3  &  2 \\ 
    $^{11}$C($\alpha$,p)      & $^{6}$Li(stable)    
       &  3.4  &  1 \\ 
    $^{11}$C($\alpha,\alpha$) & $^{3}$H($\beta^-$; 12.3y, 18.6 keV) 
       &  8.0  &  1$^{*}$ \\ \hline\hline
    $^{11}$C\{n,p\}           & $^{9}$B(p+2$\alpha$; 0.54 keV, 1.07 MeV)     
       &  7.1  &  2 \\ 
    $^{11}$C\{n,$\alpha$\}    & $^{6}$Be(2p+$\alpha$; 92 keV, 4.3 MeV) 
       &  2.5  &  2 \\
    $^{11}$C\{p,$\alpha$\}    & $^{6}$Li(stable) 
       & 37.3  &  1 \\ \hline \hline
   \end{tabular}
  \end{center}
\end{table}

The first upper block in the table just shows the branching ratios for
one-step emission into the $\gamma$, neutron, proton, and $\alpha$ channels.   
In the second block the branching of the first-step neutron channel is 
divided into the contributions directly to the ground state and the 
contributions to the excited states in the daughter nucleus $^{10}$C, 
which undergo subsequent second-step $\gamma$, neutron, proton, or $\alpha$ 
emission. In the same way, the first-step proton and $\alpha$ channels are 
split up in the lower blocks of Table~\ref{tab1}. In comparison to the 
time scale relevant for the detection of the ejectiles, the emission of 
de-excitation particles occurs simultaneously. Hence the ``symmetric'' 
channels like $^{11}$C(n,p) and $^{11}$C(p,n) will lead to identical 
signatures in the detector and therefore we list the sum of these branchings 
in the lowest block of Table~\ref{tab1}. Next to the two-step de-excitation 
mode, shown in the first column of Table~\ref{tab1}, the residual daughter 
nuclei are listed in the second column. Some of these nuclei are radioactive 
and can provide a time-correlated signal with the emitted particles in the
detection process. The third column gives branching ratios in percentages 
per one neutron in the $s_{1/2}$ state.

\begin{figure*}
\includegraphics[angle=90,width=5.5in]{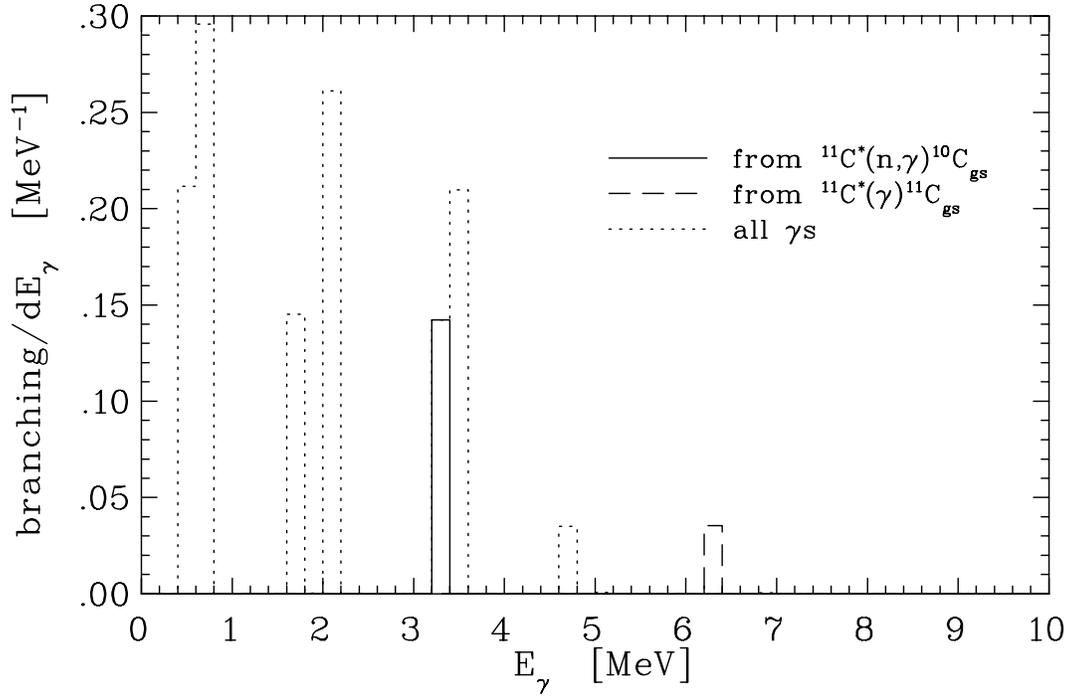}
\caption{\label{fig3}Spectra of $\gamma$s from de-excitations of 
$s_{1/2}$ hole of neutron disappearance in $^{12}$C. The dotted line is for 
all $\gamma$s from all de-excitation processes; the dashed line is for 
$^{11}$C$^{*}$($\gamma$)$^{11}$C$_{gs}$ transition, and the solid line is for 
the de-excitation mode $n + \gamma$ with a residual $^{10}$C$_{gs}$ nucleus.
The energy bin width is 0.2 MeV.}
\end{figure*}

\begin{figure*}
\includegraphics[angle=90,width=5.5in]{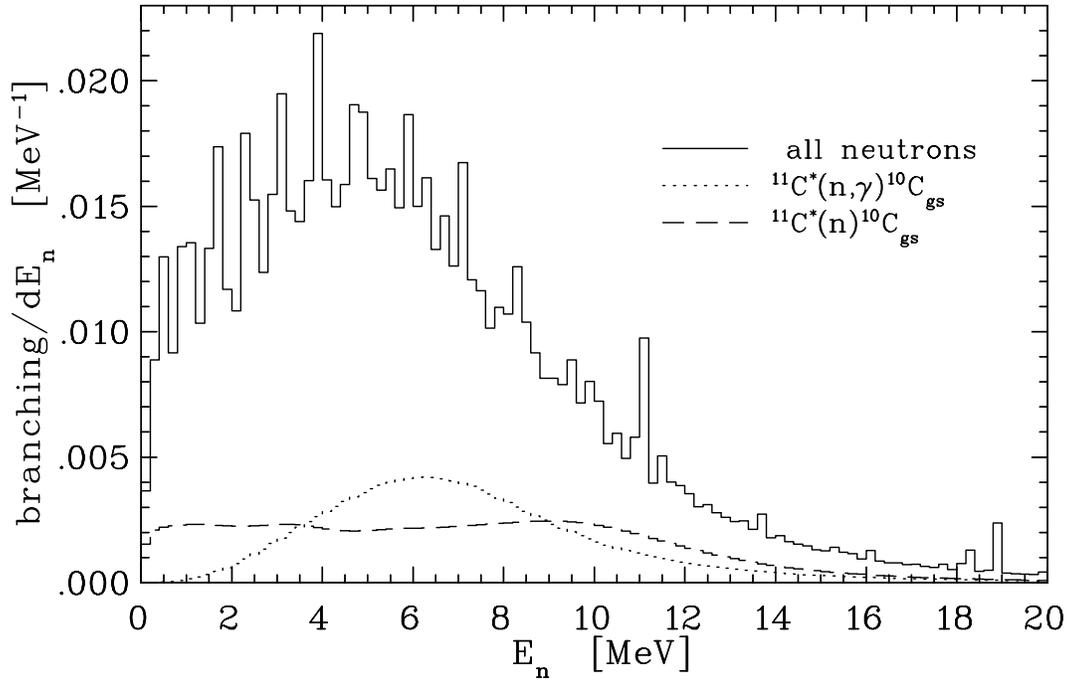}
\caption{\label{fig4}Spectra of neutrons from de-excitations of $s_{1/2}$ 
hole of neutron disappearance in $^{12}$C. The solid line is for all neutrons from
all de-excitation processes; the dotted line is for the process with emission 
of de-excitation $n$, $\gamma$, and $^{10}$C$_{gs}$ residual nucleus;
the dashed line is for the process with de-excitation via emission of neutron
and $^{10}$C$_{gs}$ residual nucleus. The energy bin width is 0.2 MeV.}
\end{figure*}

Energy spectra for de-excitation $\gamma$s and neutrons are shown in Figures 
\ref{fig3} and \ref{fig4} respectively. Figure \ref{fig3} shows the calculated 
$\gamma$-spectra for all de-excitation modes (dotted line), for 
$^{11}$C$^{*}$($\gamma$)$^{11}$C$_{gs}$ transitions (dashed line), and for 
the transition $^{11}$C$^{*}(n,\gamma)^{10}$C$_{gs}$ (solid line). The last 
spectrum is dominated by a strong monoenergetic line at 3.35~MeV, 
corresponding to the decay of the first excited $2^{+}$ state in 
$^{10}$C. Figure \ref{fig4} shows the energy spectra for the emitted 
neutrons. While the solid line depicts the summed neutron spectra 
from all processes following the disappearance of a $s_{1/2}$ neutron,
the dotted line just shows the neutron energy distribution from the 
$^{11}$C$^{*}(n,\gamma)^{10}$C$_{gs}$ mode, and the dashed line 
relates to the neutrons from the decay $^{11}$C$^{*}(n)^{10}$C$_{gs}$. 
The spectrum of neutrons from the $^{11}$C$^{*}(n,\gamma)^{10}$C$_{gs}$ 
mode reflects the width of the initial $s_{1/2}$ neutron level in $^{12}$C.

In liquid scintillator detectors the emission of de-excitation $\gamma$s, 
protons, and $\alpha$s will be seen as a prompt signal, and we assumed 
here that these particles cannot be distinguished. The neutrons, although 
emitted at the same time as other de-excitation products, will result 
in a double signal. First, they will quickly slow down by elastic collisions 
with hydrogen (which, besides $^{12}$C, is another essential component of the 
liquid scintillator) and thus will also contribute to the prompt component of 
the signal. Second, thermalized neutrons will diffuse in the liquid scintillator 
for typically 200 $\mu$sec before they are captured by hydrogen atoms with emission 
of a detectable 2.2~MeV photon. Thus, neutrons can provide two signals in the chain 
of detected events. Residual radioactive nuclei left after one- or two-particle 
de-excitations will provide another signal in the chain of events initiated 
by the disappearance of the neutron, all coming from the same point in the 
detector within the reconstruction space resolution and within certain 
correlation time.  We also note that positrons from $\beta^{+}$ decay
of the daughter nuclei in Table \ref{tab1} can be seen in the liquid 
scintillator detector with efficiency close to 100\%. 

The chain of time- and space-correlated events creates a unique signature for
the detection of neutron disappearance in low-threshold liquid scintillator 
detectors. The right-most column of Table \ref{tab1} shows the experimental 
signatures of the corresponding decay modes in terms of time- and 
space-correlated hits in the detector. One can see that the decay chains: 
$^{11}$C$^{*} \rightarrow n + \gamma + ^{10}$C$_{gs}$, with a branching of 
2.8\%, $^{11}$C$^{*} \rightarrow n + ^{10}$C$_{gs}$, with a branching of 
3.0\%, and $^{11}$C$^{*} \rightarrow n + n + ^{9}$C$_{gs}$, with a branching 
of 0.06\% provide 3-hit or 4-hit signatures. Therefore these decay modes 
should be very immune to possible accidental-coincidence background.

As the initial values for the excitation energy $E^{*}$ and spreading width
$\Gamma$ are not precisely known, in order to estimate the corresponding 
systematic stability of our results, we varied these quantities within their
uncertainty range: $\pm 1$~MeV for $E^{*}$ and from 6 MeV to 15 MeV for $\Gamma$.
Exemplary for all decay modes, the effect of the 
variations on the results are shown in Table \ref{tab2} for the branching to 
the $^{11}C^{*} \rightarrow n + \gamma + ^{10}C_{gs}$ channel, which we 
consider to deliver the most robust signature for a neutron disappearance. 

\begin{table}
   \caption{\label{tab2} Branching ratio (in percent) into the 
            $^{11}$C$^{*}$(n,$\gamma$)$^{10}$C$_{\rm gs}$
            decay channel as a function of binding energy
            and spreading width of $s_{1/2}$ neutron level in $^{12}$C.}
   \begin{center}
     \begin{tabular}{|c||c|c|c|c|c|c|} \hline
               & \multicolumn{6}{|c|}{$\Gamma_{s_{1/2}}$ [MeV]} \\ \hline
        $E_{s_{1/2}}$ [MeV]
         & 6 & 7 & 8 & 10 & 12 & 15 \\  \hline\hline
        40.7 & 2.90  &  2.76  &  2.65 & 2.46 & 2.31 & 2.14 \\ \hline
        41.7 & 2.97  &  2.84  &  2.73 & 2.54 & 2.38 & 2.20 \\ \hline
        42.7 & 2.97  &  2.85  &  2.75 & 2.57 & 2.42 & 2.24 \\ \hline
     \end{tabular}
   \end{center}
\end{table}

One can see from Table \ref{tab2} that the predicted branching ratio is 
rather insensitive to the variation of $E^{*}$ and 
$\Gamma$. Since, as mentioned above, an error of 30\% should be assigned 
to the branching ratios, we consider the systematic error coming   
from the uncertainty in $E^{*}$ and $\Gamma$ to be negligible.

In this paper we assume that decay products on neutron disappearance
(e.g.  \n3nu) have total momentum equal to the momentum of the initial 
neutron. That can be justified by the assumption that disappearance of a 
neutron occurs at very short distances and for very short times relative 
to the distances and times that are characteristic of nuclear processes. 
Therefore, one can expect that the momentum of the disappearing neutron 
is taken away by the decay products except possibly for a small recoil 
given to the residual nucleus as a whole.
Our main result, the branching ratios for the residual nucleus de-excitation, 
will not be affected by the recoil motion. However, spectra of emitted protons, 
$\alpha$s, and neutrons might be smeared out by this motion. From experimental \
point of view this smearing will not lead to any principal effects, it might only 
slightly increase the uncertainty of the efficiency of detection of these 
particles due to the finite visible energy threshold.

\section{\label{Osect}Neutron Disappearance from $^{16}$O}

Neutron disappearance from the $s_{1/2}$ state in $^{16}$O followed
by $\gamma$ de-excitation from a highly excited state has been 
considered in \cite{Suzuki93} for a Kamiokande experiment with a
water-\v{C}erenkov detector. The branching ratio for this de-excitation
mode was calculated in this paper. It is interesting to compare our 
branching ratio calculations for the initial $^{12}$C nucleus with 
this result. Integrating $\gamma$ spectrum of $^{11}$C$^{*}$ 
de-excitation, which is shown in Figure \ref{fig3} as a dotted line, 
within the energy range from 19 to 50 MeV, as suggested by 
Ref.~\cite{Suzuki93}, we found a branching ratio of 1.1$\cdot$10$^{-4}$. 
This is similar to the $s_{1/2}$ hole $\gamma$ de-excitation branching 
ratio in $^{16}$O, calculated in \cite{Suzuki93} as 
(0.27 $-$ 1.04)$\cdot$10$^{-4}$. 

We have also calculated the spectrum of final de-excitation states 
for an excited $^{15}$O nucleus, originated by a neutron hole in the
S$_{1/2}$ neutron level in $^{16}$O. These branching ratios are shown 
in Table \ref{tab3}. They were obtained by SMOKER code calculations 
assuming the values $E^{*} = 29$~MeV and $\Gamma=7$~MeV respectively 
for the energy and spreading width of the excited state in $^{15}$O.

\begin{table}
  \caption{\label{tab3}Branching ratios for $^{15}$O$^*$ de-excitation 
           after neutron disappearance from a $s_{1/2}$ hole state 
           in $^{16}$O.}
  \begin{center}
   \begin{tabular}{|l|l|r|} \hline \hline
     Decay & Daughter & Mode \\ 
     mode  & (decay, T$_{1/2}$ or $\Gamma$, Q$_{EC}$)  & \%   \\ \hline\hline
     $^{15}$O($\gamma$)        & $^{15}$O$_{\rm gs}$($\beta^+$; 122.2s, 2.75 MeV) 
       &  0.5 \\ 
    $^{15}$O(n$\ldots$)       & $\downarrow$       
       & 14.2 \\
    $^{15}$O(p$\ldots$)       & $\downarrow$       
       & 64.3 \\
    $^{15}$O($\alpha \ldots$) & $\downarrow$ 
       & 21.0 \\ \hline\hline
    $^{15}$O(n)               & $^{14}$O$_{\rm gs}$($\beta^+$; 70.6s, 5.14 MeV)
       &  1.1 \\ 
    $^{15}$O(n,$\gamma$)      & $^{14}$O$_{\rm gs}$($\beta^+$; 70.6s, 5.14 MeV)
       &  0.0 \\ 
    $^{15}$O(n,n)             & $^{13}$O($\beta^+$; 8.58ms, 17.8 MeV)     
       &  0.0 \\ 
    $^{15}$O(n,p)             & $^{13}$N($\beta^+$; 9.97s, 2.22 MeV)     
       & 12.6 \\ 
    $^{15}$O(n,$\alpha$)      & $^{10}$C($\beta^+$; 19.3s, 3.65 MeV)
       &  0.5 \\ \hline
    $^{15}$O(p)               & $^{14}$N(stable)
       &  3.4 \\ 
    $^{15}$O(p,$\gamma$)      & $^{14}$N(stable)
       & 10.4 \\ 
    $^{15}$O(p,n)             & $^{13}$N($\beta^+$; 9.97s, 2.22 MeV)
       & 11.3 \\ 
    $^{15}$O(p,p)             & $^{13}$C(stable)    
       & 30.4 \\ 
    $^{15}$O(p,$\alpha$)      & $^{10}$B(stable) 
       &  8.8 \\ \hline
    $^{15}$O($\alpha$)        & $^{11}$C($\beta^+$; 20.4m, 1.98 MeV)
       &  1.4 \\ 
    $^{15}$O($\alpha,\gamma$) & $^{11}$C($\beta^+$; 20.4m, 1.98 MeV)
       &  6.0 \\ 
    $^{15}$O($\alpha$,n)      & $^{10}$C($\beta^+$; 19.3s, 3.65 MeV)
       &  0.4 \\ 
    $^{15}$O($\alpha$,p)      & $^{10}$B(stable)    
       &  4.5 \\ 
    $^{15}$O($\alpha,\alpha$) & $^{7}$Be($\beta^+$; 53.3d, 0.86 MeV) 
       &  8.7 \\ \hline\hline
    $^{15}$O\{n,p\}           & $^{13}$N($\beta^+$; 9.97s, 2.22 MeV)
       & 23.9 \\ 
    $^{15}$O\{n,$\alpha$\}    & $^{10}$C($\beta^+$; 19.3s, 3.65 MeV)
       &  0.9 \\
    $^{15}$O\{p,$\alpha$\}    & $^{10}$B(stable)    
       & 13.3 \\ \hline \hline
   \end{tabular}
\end{center}
\end{table}

\begin{figure*}
\includegraphics[angle=90,width=5.6in,bb=40 51 516 775,clip]{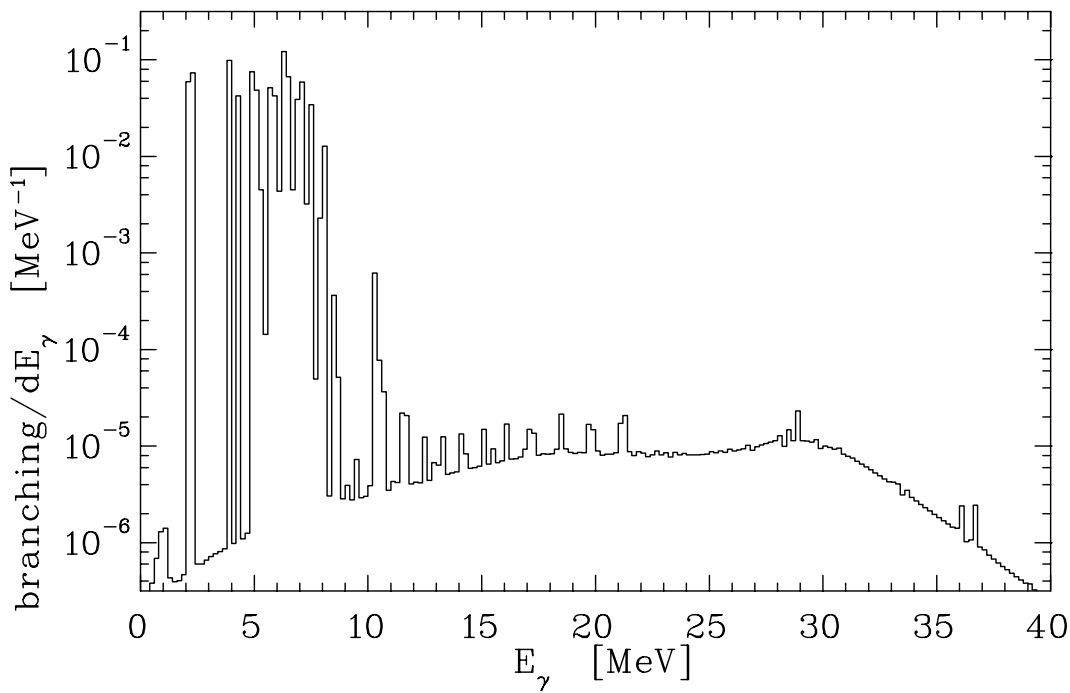}
\caption{\label{fig5}Spectrum of $\gamma$s from de-excitation 
of a $s_{1/2}$ hole resulting from neutron disappearance in 
$^{16}$O. The energy bin width is 0.2 MeV.}
\end{figure*}

Figure \ref{fig5} shows in a logarithmic scale the spectrum of 
$\gamma$s from S$_{1/2}$ de-excitations of $^{15}$O$^{*}$. 
For the $\gamma$ branching ratio, integrated over the energy 
range 19$-$50 MeV, we obtain the value 1.4$\cdot$10$^{-4}$. 
By varying the energy level and width of the $s_{1/2}$ state 
in $^{16}$O by $\pm$ 1 MeV, we found the branching to change in 
the range (1.3$-$1.5) $\cdot 10^{-4}$. Furthermore, taking 
into account the uncertainty in the threshold energy of 19 MeV 
due to the energy scale calibration of $\pm$ 0.6 MeV, the range 
becomes (1.2$-$1.6)$\cdot 10^{-4}$. Finally, estimating the 
systematic uncertainties of the SMOKER code, we conclude that 
$\gamma$ branching for de-excitation of the S$_{1/2}$ hole in 
$^{16}$O within the energy range 19$-$50 MeV is 
$(1.4\pm 0.7) \cdot 10^{-4}$, which is in good agreement with 
the result (0.27$-$1.04)$\cdot 10^{-4}$ of the calculations 
from Ref.~\cite{Suzuki93}. 

\section{\label{sect4}Two-neutron disappearance from~$^{12}$C}

Since the disappearance of two neutrons would be a process occurring at a 
very short distance, we should consider both disappearing neutrons to be 
in the same subshell, and thus we will neglect the cases when one 
S-shell and one P-shell neutrons are disappearing together.

\subsection{Disappearance from the $p_{1/2}$ or $p_{3/2}$ shell}

In the simple shell-model picture, a disappearance of two neutrons 
from the $p_{3/2}$ shell would always lead to $J^{\pi}=0^+$ 
ground state of $^{10}$C. 
However, due to correlations and particularly due to 
nucleon pairing, as was discussed earlier, there is a probability of $\sim$ 
40 $\%$ of finding a pair of neutrons in the $p_{1/2}$ level. 
We again assume that the disappearance of two neutrons would occur due 
to a very short-range interaction, and therefore expect the disappearance of 
paired neutrons in either a $p_{1/2}$ or a $p_{3/2}$ state to be more likely 
than the disappearance of a pair of one $p_{1/2}$ and one $p_{3/2}$ neutrons. 
Nevertheless, we will discuss both cases below. 

The disappearance of paired neutrons with total spin~0 would imply 
certain selection rules for the final state of disappearance. 
Thus, from the conservation of angular momentum, the transition 
$nn \rightarrow \nu \bar{\nu}$ with $\Delta (B-L)=-2$ is possible while 
transitions $nn \rightarrow \nu \nu$ or $nn \rightarrow \bar{\nu} 
\bar{\nu}$ are not. 
The disappearance of two neutrons from the $p_{1/2}$ 
level will lead to the ground state of $^{10}$C.
The disappearance of two neutrons from a $p_{3/2}$ state, while two other 
neutrons are in a $p_{1/2}$ state, might result in the excited state of 
$^{10}$C with $J^{\pi}= 0^+$. Since there are no excited states
of $^{10}$C with this spin and parity below the proton separation 
threshold, the residual de-excitation will be dominated by 
$^{10}$C$^{*} \rightarrow p + ^{9}$B, and $^{9}$B will quickly 
decay into $p + 2 \alpha$. Such de-excitation will lead only to 
a single-hit signature in the liquid scintillator detector and 
therefore is not favorable for detection.

A disappearance of one $p_{3/2}$ and one $p_{1/2}$ neutron would lead 
to an excited state in $^{10}$C with spin and parity $J^{\pi}= 2^+$ or $1^+$.  
There is not any low lying energy level known in $^{10}$C 
with spin and parity $1^+$~\cite{Aj90}.
However, there are 3 low lying energy levels stated in $^{10}$C 
(two are well established and one is assumed) with spin and parity 
$2^+$~\cite{Aj90} that could be populated after 2-neutron disappearance. 
To our knowledge, no experimental or theoretical information on the 
spectroscopic factors for these possible final states exists here. 
However, as 2 out of the three excited $2^+$ states in $^{10}$C lie 
above the proton or even $\alpha$ separation threshold, their decay will 
proceed by emission of a low energetic proton or $\alpha$-particle, leading
to the $^{9}$B and $^{6}$Be ground states, respectively. 
As $^{9}$B immediately decays into a proton and two $\alpha$ particles,
and the $^{6}$Be into 2 protons and $\alpha$, this will lead only to a
single-hit signature in the liquid scintillator detector.  
Only the population of the first excited $2^+$ state at 3.35 MeV
will produce a space and time correlated two-hit event in the detector.
This state decays by emission of a photon with energy 3.35 MeV, followed 
by the $\beta ^{+}$ decay of the residual $^{10}$C$_{gs}$ with a correlation 
time of 27.8 s and detectable energy between 1.74 and 3.65 MeV. 
As an upper limit for the probability of this event we adopt the value 
40 percent $\times$ 4/6 (probability to pick a ($p_{3/2}$,$p_{1/2}$) 
pair from two $p_{3/2}$ and two $p_{1/2}$ neutrons)  / 3 (number
of low lying $2^+$ states in $^{10}$C) or 9 \% per $^{12}$C atom, with 
an estimated error of $\pm$ 5 \%.

\subsection{Disappearance from the $s_{1/2}$ shell}

For the disappearance of two neutrons from the $s_{1/2}$ state, 
we estimated the 
excitation energy of $^{10}$C$^{*}$ in the following way: bound energy of 
two neutrons in the $s_{1/2}$ state is assumed as 78 MeV; 
separation energy of two 
neutrons from $^{12}$C is 31.8 MeV; thus the excitation energy is $\sim$46~MeV.
With respect to the spreading width of a double $s_{1/2}$ hole state one might
be inclined to assume that it should be larger than the width of a single 
$s_{1/2}$ hole (7~MeV). However, essential contribution to the spreading width 
of the single $s_{1/2}$ hole is due to the pairing correlations
that should be significantly reduced for a double $s_{1/2}$ hole state.
Therefore, a width of $\Gamma$=7~MeV has been assumed for a double $s_{1/2}$ 
hole state. The results of SMOKER code calculations for this case are shown 
in Table \ref{tab4} and were found to change their relative value by less 
than 3.5 percent when varying the spreading width between $\Gamma$=4 and 15~MeV.

\begin{table}[!b]
\caption{\label{tab4}Branching ratios for $^{10}$C$^*$ de-excitation after 
         two-neutron disappearance from $s_{1/2}$-state of $^{12}$C.}
  \begin{center}
   \begin{tabular}{|l|l|r|c|} \hline \hline
     Decay & Daughter & Mode  & Exp.   \\ 
     mode  & (decay, T$_{1/2}$ or $\Gamma$, Q$_{EC}$)  & \%    & sign.  \\
           &          &       & (hits) \\ \hline\hline
    $^{10}$C($\gamma$)        & $^{10}$C$_{\rm gs}$($\beta^+$, 19.3 s, 3.65 MeV)
       &  0.0  & - \\ 
    $^{10}$C(n$\ldots$)       & $\downarrow$       
       & 12.2  & $\downarrow$ \\
    $^{10}$C(p$\ldots$)       & $\downarrow$       
       & 63.7  & $\downarrow$ \\
    $^{10}$C($\alpha \ldots$) & $\downarrow$ 
       & 24.1  & $\downarrow$ \\ \hline\hline
    $^{10}$C(n)               & $^{9}$C($\beta^{+}$, 0.127 s, 16.5 MeV)
       &  6.2  & 3 \\ 
    $^{10}$C(n,$\gamma$)      & $^{9}$C
       &  0.0  & - \\ 
    $^{10}$C(n,n)             & $^{8}$C
       &  0.0  & - \\ 
    $^{10}$C(n,p)             & $^{8}$B($\beta^{+}\alpha$, 770 ms, 18 MeV)
       &  6.0  & 3 \\ 
    $^{10}$C(n,$\alpha$)      & $^{5}$Be
       &  0.0  & -\\ \hline
    $^{10}$C(p)               & $^{9}$B
       &  0.0  & - \\ 
    $^{10}$C(p,$\gamma$)      & $^{9}$B
       &  0.0  & - \\ 
    $^{10}$C(p,n)             & $^{8}$B
       &  0.0  & - \\ 
    $^{10}$C(p,p)             & $^{8}$Be(2$\alpha$, 6.8 eV, Q$_{\alpha}$=92 keV)
       & 27.3  & 1 \\ 
    $^{10}$C(p,$\alpha$)      & $^{5}$Li(p+$\alpha$, 1.5 MeV, Q$_{p}$=1.97 MeV)
       & 36.4  & 1 \\ \hline
    $^{10}$C($\alpha$)        & $^{6}$Be(2p+$\alpha$, 92 keV, Q$_{p\alpha}$=1.37 MeV)
       &  8.7  & 1 \\ 
    $^{10}$C($\alpha,\gamma$) & $^{6}$Be
       &  0.0  & - \\ 
    $^{10}$C($\alpha$,n)      & $^{5}$Be
       &  0.0  & - \\ 
    $^{10}$C($\alpha$,p)      & $^{5}$Li(p+$\alpha$, 1.5 MeV, Q$_{p}$=1.97 MeV)
       & 14.1  & 1 \\ 
    $^{10}$C($\alpha,\alpha$) & $^{2}$H(stable)
       &  1.3  & 1 \\ \hline\hline
    $^{10}$C\{n,p\}           & $^{8}$B($\beta^{+}\alpha$, 770 ms, 18 MeV)
       &  6.0  & 3 \\ 
    $^{10}$C\{n,$\alpha$\}    & $^{5}$Be
       &  0.0  & - \\
    $^{10}$C\{p,$\alpha$\}    & $^{5}$Li(p+$\alpha$, 1.5 MeV, Q$_{p}$=1.97 MeV)
       & 50.5  & 1 \\ \hline \hline
   \end{tabular}
\end{center}
\end{table}

\begin{figure*}
\includegraphics[angle=90,width=5.5in]{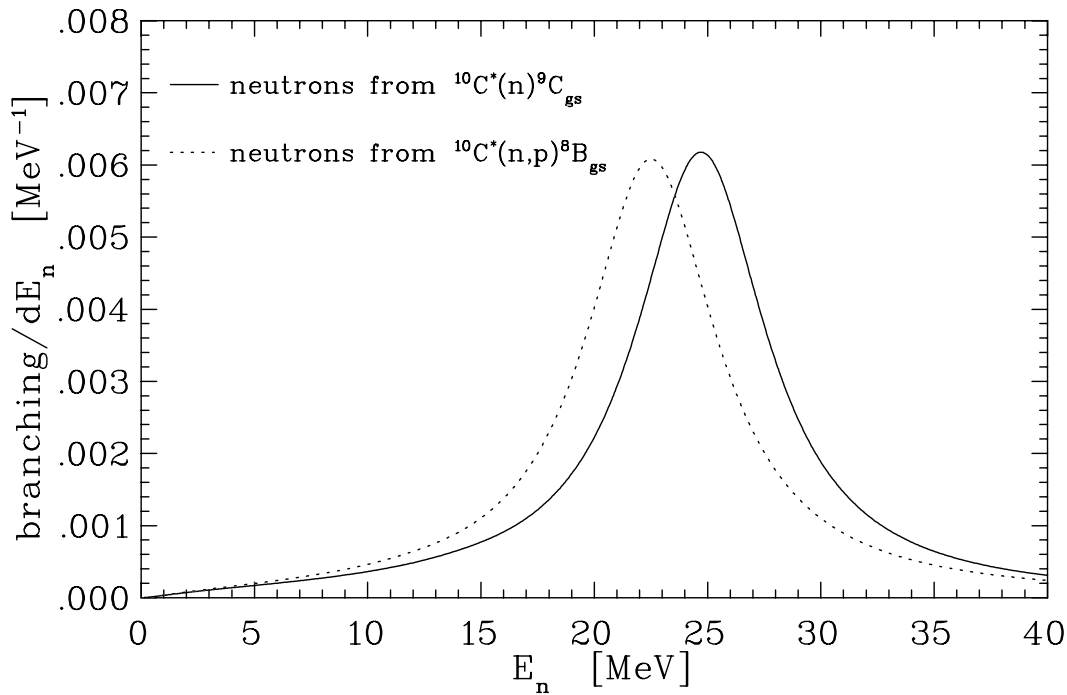}
\caption{\label{fig6}Energy spectrum of neutrons from de-excitations 
of a $s_{1/2}$ hole resulting from two-neutron disappearance in 
$^{12}$C. The solid line is for transition $^{10}$C$^{*} \rightarrow n + 
^{9}$C$_{gs}$ and the dotted line is for $^{10}$C$^{*} \rightarrow n + p 
+ ^{8}$B$_{gs}$. The energy bin width is 0.2 MeV.}
\end{figure*}

There are two final-state modes in this table that would result in a 3-hit 
event signature in the liquid scintillator detector, each with a branching of 
about 6$\%$ per S-shell neutron. Figure \ref{fig6} shows the energy spectra
of neutrons for these de-excitation modes. Note that both spectra reflect 
an initial Lorentz-shape distribution of excitation energy. For the mode 
$^{10}$C$^{*} \rightarrow n + p + ^{8}$B$_{gs}$, the proton, which comes 
from the decay of the first excited state in $^{9}$C, is monoenergetic
with energy 0.922 MeV.

\section{\label{discuss}Discussion}

The KamLAND detector \cite{Kamland} is perfectly suited to search for
the neutron disappearance processes in carbon, i.e., for the observation of
sequences of nuclear de-excitation events. The active mass of the KamLAND 
detector filled with a liquid scintillator (formula $\sim$ CH$_{2}$) 
is $\sim 900$ tons; the trigger threshold can be as low as $\sim 0.2$ MeV; 
the photon energy resolution at 1 MeV is $\leq 8$\%. Good time and energy resolution allow 
reconstruction of an event position in the detector with the accuracy down 
to several cm. Low background in the KamLAND experiment is due to a deep underground 
location (2,700 meters of water equivalent), water shield, buffer oil shield, 
veto system, radon shields, and liquid scintillator purification that provide 
concentration of $^{238}$U, $^{232}$Th, and $^{40}$K at the level below $10^{-15}$ g/g 
\cite{KamlandBackground}.

\subsection*{Experimental signatures}

We have calculated in sections \ref{sect3} and \ref{sect4} above the branching 
ratios for various de-excitations following the disappearance of one or two 
neutrons from the $^{12}$C nucleus. Not all de-excitation modes are favorable 
for detection in the experiment, but some of them form the chains of time- and 
space-correlated events in the detector. Thus, the process 
$^{11}$C$^{*} \rightarrow n + \gamma + ^{10}$C$_{gs}$ from the disappearance 
of a neutron from the $s_{1/2}$ state will look like a 3-hit event in the detector. 
The first hit in the detector will be produced by monoenergetic $\gamma$s with an 
energy of 3.35 MeV corresponding to the de-excitation of a $2^{+}$ level of $^{10}$C 
and by slow-down interactions of neutron in the scintillator. The first hit will be 
followed by a second hit, which is due to neutron capture by hydrogen in the 
scintillator (correlation time is $\sim 200 \mu$sec). The third hit will be due 
to delayed $\beta^{+}$ decay of $^{10}$C$_{gs}$ with correlation time 27.8 sec 
and detectable energy in the range 1.74--3.65 MeV. The first two hits will look 
similar to a signature of the reactor antineutrino interactions: positron plus 
delayed captured neutron. The detection of reactor antineutrinos is the main purpose 
of the KamLAND experiment \cite{Kamland}. The expected number of reactor antineutrino 
interactions in KamLAND in the absence of neutrino oscillations is $\sim 600$ per 
kt per year with the background less than few percent \cite{Kamland}. Using the measurements 
performed in the KamLAND Collaboration \cite{KamlandBackground}, one can conservatively 
estimate the background for an accidental 3-hit coincidence (first hit with a threshold 
of 2.8 MeV, second hit with a threshold of 1.6 MeV; window for first two hits is 0.54 msec; 
threshold for the third hit is 2 MeV; time window for the third hit is 60 sec; 
position-correlation cut of $\sim$ 1m) to be below one event per year. 
This accidental background rate can be measured in the detector and subtracted. 
Using branching for $^{11}$C$^{*} \rightarrow n + \gamma + ^{10}$C$_{gs}$
from Table \ref{tab1} and assuming no other sources of events with such a signature 
(e.g. from atmospheric neutrino interactions), for one year of exposition 
time in the KamLAND detector, one can, with 90\% CL, set a limit for \n3nu decay at 
the level $5 \cdot 10^{29}$ years. This is 1,000 times better than the present limit 
for this decay mode \cite{PDG2002}. 

Detection of the 3-hit process $^{11}$C$^{*} \rightarrow n + ^{10}$C$_{gs}$ will be 
similar to the one discussed above, probably with a different threshold for the first 
hit that will be formed by the slow-down interactions of neutrons in the liquid scintillator.
Detection of the 4-hit process $^{11}$C$^{*} \rightarrow n + n + ^{9}$C$_{gs}$ would
be a spectacular observation; unfortunately its branching, as follows from 
Table \ref{tab1}, is much lower than for two processes with the 3-hit signature
discussed above.

The two-hit events listed in Table \ref{tab1} should be divided into two categories.
Events with one of the de-excitation particles being a neutron belong to the first 
category; in KamLAND these events will have a signature similar to that for 
anti-neutrino reactor events with a corresponding background and will be difficult 
to disentangle. The second category should include events with radioactive daughter 
nuclei. Events resulting from the neutron disappearance from the $p_{3/2}$ state of 
$^{12}$C should also belong in this category. Accidental coincidence background 
will be more of a problem for the events with long lifetime daughter isotopes
therefore only modes with a production of $^{11}$C$_{gs}$ might have practical 
interest for the detection. Correlation time between two hits in this case is 
20.4 min due to lifetime of $^{11}$C$_{gs}$. The statistical fluctuations of the 
background will determine the sensitivity limit for a corresponding process search. 
Optimistically, this measurement alone can improve the existing lifetime limit for 
\n3nu by approximately an order of magnitude.

De-excitation final states originated by the disappearance of two neutrons from 
$^{12}$C provide two interesting 3-hit signatures: $^{10}$C$^{*} \rightarrow 
n + ^{9}$C$_{gs}$with a branching of 6.2\% and 
$^{10}$C$^{*} \rightarrow n + p + ^{8}$B with a branching of 6.0\%. 
The relatively short lifetime of the daughter nuclide makes such events free of 
accidental background. 
 
\subsection*{Non-accidental background}
 
It is interesting to see whether atmospheric neutrino events can provide 
a signature similar to $^{11}$C$^{*} \rightarrow n + \gamma + ^{10}$C$_{gs}$,
$^{11}$C$^{*} \rightarrow n + ^{10}$C$_{gs}$, or other processes. We anticipate 
in the future to study this interesting question in a separate paper treating 
nuclear de-excitations produced by atmospheric neutrinos and anti-neutrinos in 
$^{12}$C. Such a study will also be interesting from the point of view of 
identification of atmospheric neutrino and anti-neutrino interactions in  
liquid scintillator detectors through the detection of associated nuclear 
chains of de-excitations and radio-nuclide decays.
In the respect of the background for one-neutron or two-neutron disappearance 
addressed in this paper, few qualitative comments can be made here. Charged
current atmospheric neutrino reactions have a rate of about 100 interactions 
per kt-detector per year. Due to the relatively high energy spectrum of atmospheric 
neutrinos, leptons produced in charged current interactions usually have an energy 
larger than the energy typical for nucleus de-excitation. We remind the reader
that in the Kamiokande experiment \cite{Suzuki93} the energy range between 
19$-$50 MeV was practically free of background. Thus, one can hope that charge 
current processes initiated by atmospheric neutrinos will not produce significant 
background for neutron disappearance processes, i.e., the chain of events with 
the first signal in a few-MeV range. However, detailed simulations of atmospheric 
neutrino interactions with $^{12}$C using the SMOKER code for treatment of nuclear 
de-excitations are needed to understand this background quantitatively. In such 
simulations, the production of low energy charged leptons has to be estimated together 
with de-excitation of $^{12}$N$^{*}$ and $^{12}$B$^{*}$ nuclei with a continuous 
spectrum of excitations. 
Neutral current atmospheric neutrino interactions should be also followed by 
de-excitations of $^{12}$C$^{*}$ (see paper \cite{supernova}) and might, for 
example, potentially lead to the process $^{12}$C$^{*} \rightarrow 
n + n + \gamma + ^{10}$C$_{gs}$ with one of the neutrons lost at the edge 
of the fiducial volume of the detector. 

\section{\label{conclude}Conclusions}

In this paper we show that modern high-mass, low-threshold, and low-background
liquid scintillator detectors can observe unique signatures of chains of time- 
and space-correlated events, formed by nuclear de-excitations following the 
neutron or two-neutron disappearance from the nucleus. The KamLAND detector 
\cite{Kamland} is perfectly suited to search for such processes originating from 
$^{12}$C nuclei. The sensitivity of search for a neutron disappearance
(e.g. \n3nu) or two-neutron disappearance (e.g. $nn \rightarrow \nu \bar{\nu}$)
in the KamLAND detector can be extended by 3-4 orders of magnitude relative to 
the present lifetime limits for these processes. Non-accidental background for 
neutron and two-neutron disappearance from atmospheric neutrino interactions, 
although expected to be small, needs to be calculated in future work.  
Assuming that nucleons can decay only into known particles with conservation 
of electric charge, 4-momentum, and angular momentum, the search for intra-nucleus 
neutron and two-neutron disappearance in KamLAND can lead either to the discovery 
of long-awaited baryon instability or to the improvement of the mode-independent 
nucleon decay lifetime limit from the present 1.6$\cdot$10$^{25}$ years to or 
above $\sim$10$^{30}$ years.

We think that the calculations performed in this paper with regard to the KamLAND 
detector and $^{12}$C nuclei can be extended to other high-mass, low-threshold,
and low-background detectors using different nuclei. We plan to continue 
exploring these possibilities.

\section*{ACKNOWLEDGMENTS}

We thank F.-K. Thielemann for providing us with the SMOKER code.
We are grateful to Petr Vogel, Andreas Piepke, and Robert Svoboda 
for useful discussions and critical remarks. Oak Ridge National 
Laboratory is managed by UT-Battelle, LLC for the U.S. Department 
of Energy under contract DE-AC05-00OR22725.


\begin{thebibliography}{50}

\bibitem{PDG2002}K.~Hagiwara {\it et al.}  
[Particle Data Group Collaboration],``Review Of Particle Physics,''
Phys.\ Rev.\ D {\bf 66}, 010001 (2002).

\bibitem{Kamland}
P.~Alivisatos {\it et al.},``KamLAND: A liquid scintillator 
anti-neutrino detector at the  Kamioka site,''
STANFORD-HEP-98-03, Tohoku-RCNS-98-15; K.~Eguchi {\it et al.}  
[KamLAND Collaboration],``First results from KamLAND: 
Evidence for reactor anti-neutrino  disappearance,''
Phys.\ Rev.\ Lett.\  {\bf 90}, 021802 (2003).

\bibitem{Sakharov67} A.~D.~Sakharov,``Violation of CP Invariance, 
C Asymmetry, and Baryon Asymmetry of the Universe,''
Pisma Zh.\ Eksp.\ Teor.\ Fiz.\  {\bf 5}, 32 (1967)
[JETP Lett.\  {\bf 5}, 24 (1967)].

\bibitem{Salam73}J.~C.~Pati and A.~Salam,``Unified Lepton - Hadron 
Symmetry And A Gauge Theory of The Basic Interactions,''  
Phys.\ Rev.\ {\bf D8}, 1240 (1973); ``Is Baryon Number Conserved?,''
Phys.\ Rev.\ Lett.\ {\bf 31}, 661 (1973);``Lepton Number as the Fourth Color,''
Phys.\ Rev.\ {\bf D10}, 275 (1974).

\bibitem{Georgi74}H.~Georgi and S.~L.~Glashow,``Unity Of All Elementary 
Particle Forces,''Phys.\ Rev.\ Lett.\  {\bf 32}, 438 (1974).

\bibitem{Goldhaber96}M.~Goldhaber, ``Search for Nucleon Instability (Origin 
and History), in {\it Proceedings of International Workshop on Future Prospects 
of Baryon Instability Search in p-Decay and \nnbari Oscillation Experiments}, 
Oak Ridge, Tennessee, pp~1-6 (1996).

\bibitem{Babu2000}K.~S.~Babu, J.~C.~Pati and F.~Wilczek,
``Fermion masses, neutrino oscillations, and proton decay in the light of 
SuperKamiokande,'' Nucl.\ Phys.\ {\bf B566}, 33 (2000).

\bibitem{Mohapatra2001}K.~S.~Babu and R.~N.~Mohapatra,
``Observable neutron anti-neutron oscillations in seesaw models of  neutrino mass,''
Phys.\ Lett.\ {\bf B518}, 269 (2001).

\bibitem{SuperK}Y.~Hayato {\it et al.} [SuperKamiokande Collaboration],
``Search for proton decay through $p \rightarrow \bar{\nu} K^{+}$ in a large water  
\v{C}erenkov detector,'' Phys.\ Rev.\ Lett.\  {\bf 83}, 1529 (1999);~M.~Shiozawa {\it et al.} 
[SuperKamiokande Collaboration],~``Search for proton decay via $p \rightarrow e^{+} \pi^{0}$
in a large water \v{C}erenkov  detector,'' Phys.\ Rev.\ Lett.\  {\bf 81}, 3319 (1998).

\bibitem{SuperK01}M.~Smy, ``Super Kamiokande,'' invited talk at NNN'01 Workshop, 
Louisiana State University, Baton Rouge, December 10-12, 2001.

\bibitem{Evans77}R.~I.~Steinberg and J.~C.~Evans, ``Nucleon Stability: A Geochemical 
Test Independent of Decay Mode,''Science {\bf 197}, 989 (1977).

\bibitem{Pati84}J.~C.~Pati, ``Nucleon Decays Into Lepton + Lepton + Anti-Lepton + 
Mesons Within SU(4) Of Color,'' Phys.\ Rev.\ {\bf D29}, 1549 (1984).

\bibitem{Yuri99}Yu.~Kamyshkov, ``Nucleon instability and (B-L) non-conservation,'' 
in {\it Proceedings of Stony Brook 1999 Workshop on Next Generation Nucleon Decay and 
Neutrino Detector}, pp 84-87 (1999). 

\bibitem{Feinberg78}G.~Feinberg, M.~Goldhaber and G.~Steigman,
``Multiplicative Baryon Number Conservation And The Oscillation Of Hydrogen Into 
Anti-Hydrogen,'' Phys.\ Rev.\ {\bf D18}, 1602 (1978).

\bibitem{Totsuka86}Y.~Totsuka, in {\it Proceedings of the 7th Workshop on Grand 
Unification/ICOBAN '86}, edited by J.~Arafune (World Scientific, Singapore, 1986), p. 118.

\bibitem{Ejiri93}H.~Ejiri, ``Nuclear Deexcitations Of Nucleon Holes Associated With 
Nucleon Decays In Nuclei,'' Phys.\ Rev.\ C {\bf 48}, 1442 (1993).

\bibitem{Suzuki93}Y.~Suzuki {\it et al.}  [Kamiokande Collaboration],
``Study of invisible nucleon decay, \n3nu, and a forbidden nuclear transition 
in the Kamiokande detector,'' Phys.\ Lett.\ {\bf B311}, 357 (1993).

\bibitem{Dama00}R.~Bernabei {\it et al.}, ``Search For The Nucleon And Di-Nucleon 
Decay Into Invisible Channels,'' Phys.\ Lett.\ {\bf B493}, 12 (2000).

\bibitem{Learned79}J.~Learned, F.~Reines and A.~Soni, ``Limits On Nonconservation 
Of Baryon Number,'' Phys.\ Rev.\ Lett.\  {\bf 43}, 907 (1979); [Erratum-ibid.\  
{\bf 43}, 1626 (1979)].

\bibitem{Berger91}C.~Berger {\it et al.}  [Frejus Collaboration],
``Lifetime limits on (B-L) violating nucleon decay and dinucleon decay modes from 
the Frejus experiment,'' Phys.\ Lett.\ B {\bf 269}, 227 (1991).

\bibitem{Glicenstein97}J.~F.~Glicenstein,``New limits on nucleon decay modes 
to neutrinos,'' Phys.\ Lett.\ B {\bf 411}, 326 (1997).

\bibitem{Borexino}C.~Arpesella et al., Borexino Proposal, Vol 1 and 2, 
ed. G.~Bellini and R.~S.~Raghavan, Univ. of Milan, (1991). 

\bibitem{ShellMods} N. Auerbach, N. Van Giai, and O.K. Vorov,
                    Phys.\ Rev.\ C {\bf 56} R2368 (1997);
                    E. Kolbe, K. Langanke, and P. Vogel,
                    Nucl. Phys. {\bf A652}, 91-100 (1999);
David~Dean, Oak Ridge National Laboratory, Private Communication.

\bibitem{Aj90}F. Ajzenberg-Selove, Nucl.\ Phys.\ {\bf A506} 1 (1990).

\bibitem{C12eep}
L.~Lapikas, G.~van der Steenhoven, L.~Frankfurt, M.~Strikman and M.~Zhalov,
``The transparency of C-12 for protons,''
Phys.\ Rev.\ C {\bf 61}, 064325 (2000).

\bibitem{Muether_private}H.~Muether, Private communication; see also
H.~Muether, A.~Polls, Prog.\ Part.\ Nucl.\ Phys.\ {\bf 45}, 243 (2000).

\bibitem{StatModel} H. Feshbach, Theoretical Nuclear Physics, John Willey $\&$ Sons, 1990. 

\bibitem{Smoker}J.~J.~Cowan, F.~-K.~Thielemann, J.~W.~Truran, 
``The R-Processes and Nucleochronology,'' Phys.\ Rep.\ {\bf 208}, 267-394 (1991).

\bibitem{KamlandBackground}L.~Hoffman, A.~Piepke, R.~McKeown, B.~Tipton, P.~Vogel,
``Background estimates for KamLAND from Natural Radioactivity of the Detector,''
Internal report of KamLAND Collaboration, June 19, 2001, 
http://citnp.caltech.edu/kamland/radioassay/rates/; see also, F. Suekane (for KamLAND 
Collaboration), in {\it Proceedings of the Conference "Beyond the Desert 2002"}, Finland, 
Oulu, June 3-7, 2002.

\bibitem{mythesis}E.~Kolbe, ``Untersuchungen zur inelastischen Neutrinostreuung 
an Kernen und Anwendungen in der Kern- und Astrophysik,'' thesis, 
Universit\"{a}t M\"{u}nster (1992).
 
\bibitem{supernova}K.~Langanke, P.~Vogel, E.~Kolbe, ``Signal for supernova 
$\nu_\mu$ and $\nu_\tau$ neutrinos in water \v{C}erenkov detectors,''
Phys.\ Rev.\ Lett.\ {\bf 76}, 2629 (1996).

\end{thebibliography}
\end{document}